\documentclass[12pt]{article}
\usepackage{graphicx}

\def\Re{{\cal R \mskip-4mu \lower.1ex \hbox{\it e}\,}}
\def\Im{{\cal I \mskip-5mu \lower.1ex \hbox{\it m}\,}}
\def\ie{{\it i.e.}}
\def\eg{{\it e.g.}}

\def\etal{{\it et al.}}

\def\sub#1{_{\lower.25ex\hbox{$\scriptstyle#1$}}}
\def\tev{\,{\ifmmode\mathrm {TeV}\else TeV\fi}}
\def\gev{\,{\ifmmode\mathrm {GeV}\else GeV\fi}}
\def\mev{\,{\ifmmode\mathrm {MeV}\else MeV\fi}}
\def\mpl{\ifmmode M_{pl}\else $M_{pl}$\fi}
\def\mpl{\ifmmode \overline M_{Pl}\else $\bar M_{Pl}$\fi}
\def\to{\rightarrow}

\def\subw{_{\rm w}}
\def\mh{\ifmmode m\sbl H \else $m\sbl H$\fi}
\def\mch{\ifmmode m_{H^\pm} \else $m_{H^\pm}$\fi}
\def\mt{\ifmmode m_t\else $m_t$\fi}
\def\mc{\ifmmode m_c\else $m_c$\fi}
\def\mz{\ifmmode M_Z\else $M_Z$\fi}
\def\mw{\ifmmode M_W\else $M_W$\fi}
\def\mws{\ifmmode M_W^2 \else $M_W^2$\fi}
\def\mhs{\ifmmode m_H^2 \else $m_H^2$\fi}   
\def\mzs{\ifmmode M_Z^2 \else $M_Z^2$\fi}
\def\mts{\ifmmode m_t^2 \else $m_t^2$\fi}
\def\mcs{\ifmmode m_c^2 \else $m_c^2$\fi}
\def\mchs{\ifmmode m_{H^\pm}^2 \else $m_{H^\pm}^2$\fi}
\def\ztwo{\ifmmode Z_2\else $Z_2$\fi}
\def\zone{\ifmmode Z_1\else $Z_1$\fi}
\def\mtwo{\ifmmode M_2\else $M_2$\fi}
\def\mone{\ifmmode M_1\else $M_1$\fi}
\def\tb{\ifmmode \tan\beta \else $\tan\beta$\fi}
\def\xw{\ifmmode x\subw\else $x\subw$\fi}
\def\ch{\ifmmode H^\pm \else $H^\pm$\fi}
\def\lum{\ifmmode {\cal L}\else ${\cal L}$\fi}
\def\inpb{\,{\ifmmode {\mathrm {pb}}^{-1}\else ${\mathrm {pb}}^{-1}$\fi}}
\def\infb{\,{\ifmmode {\mathrm {fb}}^{-1}\else ${\mathrm {fb}}^{-1}$\fi}}
\def\epem{\ifmmode e^+e^-\else $e^+e^-$\fi}
\def\ppb{\ifmmode \bar pp\else $\bar pp$\fi}
\def\bsg{\ifmmode B\to X_s\gamma\else $B\to X_s\gamma$\fi}
\def\bsll{\ifmmode B\to X_s\ell^+\ell^-\else $B\to X_s\ell^+\ell^-$\fi}
\def\bstt{\ifmmode B\to X_s\tau^+\tau^-\else $B\to X_s\tau^+\tau^-$\fi}
\def\lamt{\ifmmode \tilde\lambda\else $\tilde\lambda$\fi}
\def\shat{\ifmmode \hat s\else $\hat s$\fi}
\def\that{\ifmmode \hat t\else $\hat t$\fi}
\def\uhat{\ifmmode \hat u\else $\hat u$\fi}

\newskip\zatskip \zatskip=0pt plus0pt minus0pt
\def\matth{\mathsurround=0pt}
\def\lsim{\mathrel{\mathpalette\atversim<}}

\def\atversim#1#2{\lower0.7ex\vbox{\baselineskip\zatskip\lineskip\zatskip
  \lineskiplimit 0pt\ialign{$\matth#1\hfil##\hfil$\crcr#2\crcr\sim\crcr}}}

\def\grtsim{\,\,\rlap{\raise 3pt\hbox{$>$}}{\lower 3pt\hbox{$\sim$}}\,\,}
\def\lsim{\,\,\rlap{\raise 3pt\hbox{$<$}}{\lower 3pt\hbox{$\sim$}}\,\,}

\renewcommand{\thefootnote}{\fnsymbol{footnote}}

\begin{document} \begin{titlepage}
\rightline{\vbox{\halign{&#\hfil\cr
&SLAC-PUB-14206\cr
}}}
\begin{center}
\thispagestyle{empty} \flushbottom { {\Large\bf Lorentz 
Violation in Warped Extra Dimensions 
\footnote{Work supported in part
by the Department of Energy, Contract DE-AC02-76SF00515}
\footnote{e-mail:
rizzo@slac.stanford.edu}}}
\medskip
\end{center}

\centerline{Thomas G. Rizzo}
\vspace{8pt} 
\centerline{\it Stanford Linear
Accelerator Center, 2575 Sand Hill Rd., Menlo Park, CA, 94025}

\vspace*{0.3cm}

\begin{abstract}
Higher dimensional theories which address some of the problematic issues of the Standard Model(SM)  
naturally involve some form of $D=4+n$-dimensional Lorentz invariance violation (LIV). In such 
models the fundamental physics which leads to, \eg, field localization, orbifolding, the existence 
of brane terms and the compactification process all can introduce LIV in the higher dimensional theory 
while still preserving 4-d Lorentz invariance. In this paper, attempting to capture some of this 
physics, we extend our previous analysis of LIV in 5-d UED-type models to those with 5-d warped extra 
dimensions. To be specific, we employ the 5-d analog of the SM Extension of Kostelecky \etal ~which   
incorporates a complete set of operators arising from spontaneous LIV. We show that while the 
response of the bulk scalar, fermion and gauge fields to the addition of LIV operators in warped 
models is qualitatively similar to what happens in the flat 5-d UED case, the gravity sector of these 
models reacts very differently than in flat space. Specifically, we show that LIV in this warped case 
leads to a non-zero bulk mass for the 5-d graviton and so the would-be zero mode, which we identify as 
the usual 4-d graviton, must necessarily become massive. The origin of this mass term is the simultaneous 
existence of the constant non-zero $AdS_5$ curvature and the loss of general co-ordinate invariance via 
LIV in the 5-d theory. Thus warped 5-d models with LIV in the gravity sector are not phenomenologically 
viable.  
\end{abstract}



\renewcommand{\thefootnote}{\arabic{footnote}} \end{titlepage} 

%
%
%
%
%

\section{Introduction}

The possibility of 4-d Lorentz invariance violation (LIV) due to new high scale physics has become 
a common topic for discussion in recent years{\cite {livrev}} specifically within the context of theories 
of quantum gravity. In such theories it is possible for LIV to manifest itself in many ways, \eg, ($i$) 
particles of various spins and flavors may have different dispersion relations correlating their mass, energy 
and momentum{\cite {speeds}} or ($ii$) the general Poincare algebra itself may be deformed as in Doubly Special 
Relativity{\cite {dsr}} such that the Planck mass is left as an invariant quantity. One way of generating LIV 
is through the vacuum expectation value (vev) of a tensor, usually a 4-vector, which points in a 
specific direction in a preferred reference frame, \ie, via spontaneous symmetry breaking{\cite {spon}}. If 
one imagines that maintaining rotational invariance is more sacrosanct than is boost invariance then the vev may 
take the time-like form $\sim (v,0,0,0)$ in such a frame. Within this context and assuming only the field 
content of the Standard Model (SM), the SM Extension (SME) of Kostelecky and co-workers{\cite {Kost}} 
provides us with a relatively short list for the complete set of lowest-dimension, gauge invariant LIV 
operators in 4-d which may also be CPT violating. The SME can thus provide an almost model-independent 
framework in which to examine the possibility of LIV within this specific context. Of course, LIV in 4-d has 
yet to be observed and rather stringent constraints already exist on its possible nature and, hence, the 
4-d SME parameters. 

Over the past dozen years, models with extra spatial dimensions{\cite {edrev}} with effective compactification 
radii near the TeV scale 
have become a popular and reasonably successful way to address some of the outstanding issues of the 
SM such as the gauge hierarchy and flavor problems{\cite {edrev}}. From the $D=4+n$-dimensional perspective, 
though the set of actions employed in the construction of such models generally appear to be $D$-dimensionally 
Poincare invariant, we know, due to the fact that the extra dimensions are compactified and are likely 
orbifolded, that this symmetry must be broken at high scales. These compact extra $n$-dimensions are of 
a finite extent and thus do not behave as do the conventional dimensions of `infinite' extent of 4-d and 
must be in some ways `special'. While the physics of the compactification process is not directly addressed in 
these bottom-up model constructs, it is clear that any boosts which mix the 4-d co-ordinates with those of 
the $n$ compactified dimensions will correspond to broken generators and thus be LIV from the 
general $D$-dimensional point of view. Furthermore, it it quite common in such theories to have fields 
which are localized at various points in the compactified manifold or that have special, localized 
interactions, such as `brane terms', which contribute to the total action. While such theories may be 
manifestly 4-d Lorentz invariant by construction, in the full $D$-dimensions LIV is certainly present.  

Recently we have considered the possibility of explicit LIV in the case of 5-d with one extra `flat' 
dimension within the context of Universal Extra Dimensions(UED){\cite {UED}} by employing a 5-d 
generalization of the SME{\cite {flatme}}. Out of all of the potential LIV operators in the 5-d SME, 
only the presence of LIV kinetic terms were found to be `allowed', after employing field redefinitions,  
if we required that Lorentz invariance in 4-d remain intact. In that analysis,  we found that the 
addition of such kinetic terms to the action for scalar, fermion and gauge fields can lead to a number of 
interesting phenomena: ($i$) the mass spectra of the KK fields for particles of different spins and 
chiralities can be rescaled by independent overall factors. In a way, this can be viewed as the various fields 
`seeing' different values for the (constant) 55-component of the metric tensor. ($ii$) Since the 
Kaluza-Klein(KK) excitations of LH- and RH-fermions within the SM need no longer be degenerate this 
can lead to loop-induced parity violating operators, such as anapole moments, in sectors of the theory 
which were previously found to be parity conserving; ($iii$) The $Z_2$ KK-parity symmetry of UED can 
now be violated rendering, \eg, the LKP unstable through the presence of LIV and CPT-violating terms.  
This also leads to mixing among the various KK levels of a given SM particle making the excitation 
spectrum more complex.

In this paper we will extend our previous analysis to the case of LIV in warped extra dimensions, \ie, 
to the 5-d Randall-Sundrum (RS) Model{\cite {RS}} with the SM fields living in the bulk as can be 
realized on an effectively $S^1/Z_2$-orbifolded interval{\cite {nohiggs}} without brane tensions. This 
later assumption will play no role in our examination of the consequences for scalar, fermion or gauge 
fields but will simplify our analysis for the case of gravitons. In this more general case with a warped 
metric, LIV actually corresponds to a specialized form of the loss of general co-ordinate invariance which 
is expected to hold in General Relativity. This leads to important implications in the gravity sector that 
we will see below. 

Note that since we will be following the SME approach below as in our previous analysis, if the UV 
breaking of LIV is spontaneous, we can imagine that it can be described by 
a 5-vector which takes on a vev of the form $u^A\sim (0,0,0,0,1)$ in the preferred 5-d frame in which our 
calculations are performed as was done in our earlier work. This special form is chosen to leave our ordinary 
4-d space Lorentz invariant. As we will see below, the effects of LIV on the SM scalar, fermion and gauge 
fields living in the bulk of the RS model will be qualitatively similar to that found earlier in the 
case of UED and thus leads to similar new phenomenological effects as discussed above. However, the corresponding 
effects of LIV in the gravity sector are found to much more drastic since general co-ordinate invariance only 
remains in our 4-d subspace and will be quite different than that found for the flat 5-d scenario. In particular, 
we find that the 5-d graviton field now acquires a non-zero bulk mass in the $AdS_5$ background which is proportional 
to the RS curvature parameter, $k$. This in turn prevents the existence of a massless KK zero-mode that we would 
conventionally identify as the 4-d graviton. From this result we can conclude the the combination of LIV in 
the gravity sector of a higher dimensional theory living in a space of non-zero curvature cannot 
yield a phenomenologically viable model.

\section{LIV With SM Fields in the RS Bulk}

In our previous analysis of 5-d LIV in UED based on the SME approach, we found that the gauge 
invariant and CPT-conserving operators of lowest dimension took the form of modifications to the kinetic 
terms of the scalar, fermion and gauge fields of the SM. These same operators will now be of immediate 
relevance in the corresponding warped scenario. In the analysis below, we will follow the setup of the RS 
model, constructed on an interval, as is done, \eg, in Higgsless models of electroweak symmetry 
breaking{\cite {nohiggs}}. To set notation we remind ourselves that the metric on this space is given by  
\begin{equation}
ds^2= e^{-2ky}\eta_{\mu\nu}dx^\mu dx^\nu-dy^2\,,
\end{equation}
which will be defined on the interval $0\leq y\leq \pi r_c$ where $r_c$ is the corresponding compactification 
radius. Here $\eta_{\mu\nu}=diag(1,-1,-1,-1)$ is the usual Minkowski metric and $k \lsim M$ describes the curvature 
of the space, with $M$ being the 5-d Planck scale such that 4-d (reduced) Planck scale is given by $\mpl^2=M^3/k$. 
Note that in the original RS model the SM fields were localized to the $y=\pi r_c$ orbifold fixed point, \ie, 
the TeV brane, whereas here we are considering only bulk SM fields.   

We proceed by direct analogy to our earlier flat space analysis. 
For simplicity we first examine the case of a free scalar field and then follow essentially the same approach for 
fermions and gauge bosons so that the details can be omitted. The scalar action including the contribution of 
the LIV kinetic term is found to be given by
\begin{equation}
\int d^4x ~dy~\sqrt{-g} ~\Big[(\partial_A \Phi)^\dagger (\partial^A \Phi)
-\mu^2 \Phi^\dagger \Phi-k_\Phi(\partial_5 \Phi)^\dagger (\partial_5 \Phi)
\Big]\,, 
\end{equation}
where we have allowed for a standard bulk mass, $\mu$, and the dimensionless parameter $k_\Phi$ sets the 
$\sim O(1)$ coupling strength for the LIV kinetic term for the scalar. Note that apart from a rescaling  
the field $\Phi$,  
in the  presence of the LIV term, sees an effective $g_{55}=1+k_\Phi$. We thus suspect that $k_\Phi>-1$ 
will be necessary to prevent $\Phi$ from becoming tachyonic and we will see this is indeed a correct 
expectation explicitly below. Performing the Kaluza-Klein(KK) decomposition as usual 
\begin{equation}
\Phi=\sum_n \phi_n(x) \chi_n(y)\,,
\end{equation}
then integrating by-parts and orthonormalizing the fields such that 
\begin{equation}
\int ~dy ~e^{-2ky}~\chi_n(y) \chi_m(y)=\delta_{nm}\,,
\end{equation}
we obtain from the above action the following equation of motion for the 5-d wavefunctions $\chi_n(y)$ of 
the KK scalars:
\begin{equation}
-(1+k_\Phi)\partial_y \Big[e^{-4ky}\partial_y \chi_n\Big]+\mu^2e^{-4ky}\chi_n=m_n^2e^{-2ky}\chi_n\,,
\end{equation}
where the $m_n$ are the KK masses of the scalar tower states. Imposing Neumann-like boundary conditions, 
$\partial_y \chi_n(y)=0$ at $y=0,\pi r_c$, so that one obtains the conventional RS result in the limit 
$k_\Phi \to 0$, we see that 
this is just the equation for a 5-d scalar{\cite {GW}} with a bulk mass $m^2=\mu^2/(1+k_\Phi)$ and 
with the eigenvalues corresponding to the masses $\tilde m_n^2=m_n^2/(1+k_\Phi)$. The corresponding 
eigenfunctions are then given by the well-known expressions 
\begin{equation}
\chi_n(y)={{e^{2ky}}\over {N_n}} \Bigg[J_\nu(\tilde m_n/k ~e^{ky})+ b_{n\nu} Y_\nu(\tilde m_n/k ~e^{ky})
\Bigg]\,,
\end{equation}
with $J_\nu,Y_\nu$ being the familiar Bessel functions of order $\nu$ which now explicitly depends on 
the value of the parameter $k_\Phi$ as,  
\begin{equation}
\nu=\Bigg[4+{{(\mu^2/k^2)}\over {(1+k_\Phi)}}\Bigg]^{1/2}\,,
\end{equation}
and with $N_n$ being a normalization factor. We note that for practical calculations the $b_{n\nu}$ can 
be set to zero as is usual as 
can be seen by evaluating them directly from the boundary condition $\partial_y \chi_n(y)=0$ at $y=0$.  
The KK masses for the scalars are then given by the $y=\pi r_c$ boundary condition and can be expressed 
as $m_n=k\epsilon x_n\sqrt{1+k_\Phi}$ where $\epsilon=e^{-\pi kr_c}$ is the usual warp factor and where 
the $x_n$ are the roots of the familiar equation  
\begin{equation}
2J_\nu(x_n)+x_nJ'_\nu(x_n)=0\,.
\end{equation}
Here a prime denotes differentiation with respect to the co-ordinate $y$. Note that, as in the earlier 
discussed flat UED case, 
we must require that $k_\Phi>-1$ to prevent tachyonic solutions as expected above. 

One very important feature of the above equation of motion for the scalar KK excitations that we should remember 
for the gravity analysis below is that a non-zero value for the parameter $\mu$, no matter how small, prevents the 
existence of a massless zero mode being present in the spectrum{\cite {GW}}. For $\mu \sim k$ the lightest state 
is found to have a $\sim $ TeV mass.

\begin{figure}[htbp]
\centerline{
\includegraphics[width=8.5cm,angle=90]{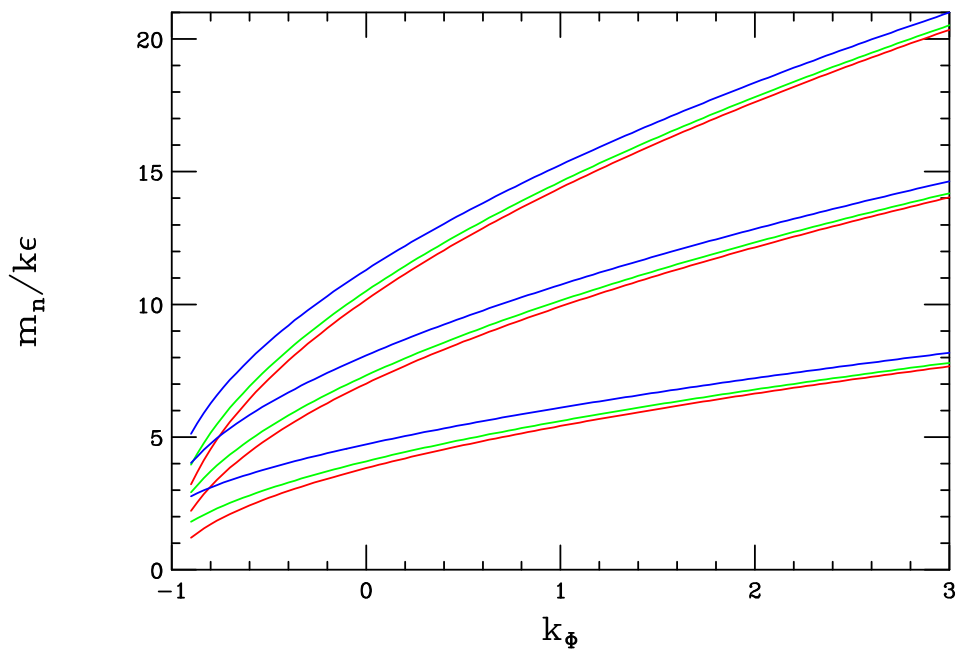}}
\vspace*{0.1cm}
\caption{Masses for the first three massive KK scalar excitations as a function of the parameter $k_\Phi$ for, 
from bottom to top within each clustering, $\mu/k=0,1,2$, respectively.}
\label{fig1}
\end{figure}
Fig.~\ref{fig1} shows the first three KK scalar tower masses as a function of the LIV parameter 
$k_\Phi$ assuming $\mu/k=0,1$ or 2. Note that the general KK spectrum is now somewhat different than is 
usual for bulk scalars; for large $k_\Phi$ the KK masses grow $\sim \sqrt {1+k_\Phi}$ which is essentially 
the same behavior as was found earlier in the flat UED scalar case. Thus as expected, although we are now dealing with 
scalar fields in a warped background the qualitative features found for scalars with LIV in flat UED model remain valid.

\begin{figure}[htbp]
\centerline{
\includegraphics[width=8.5cm,angle=90]{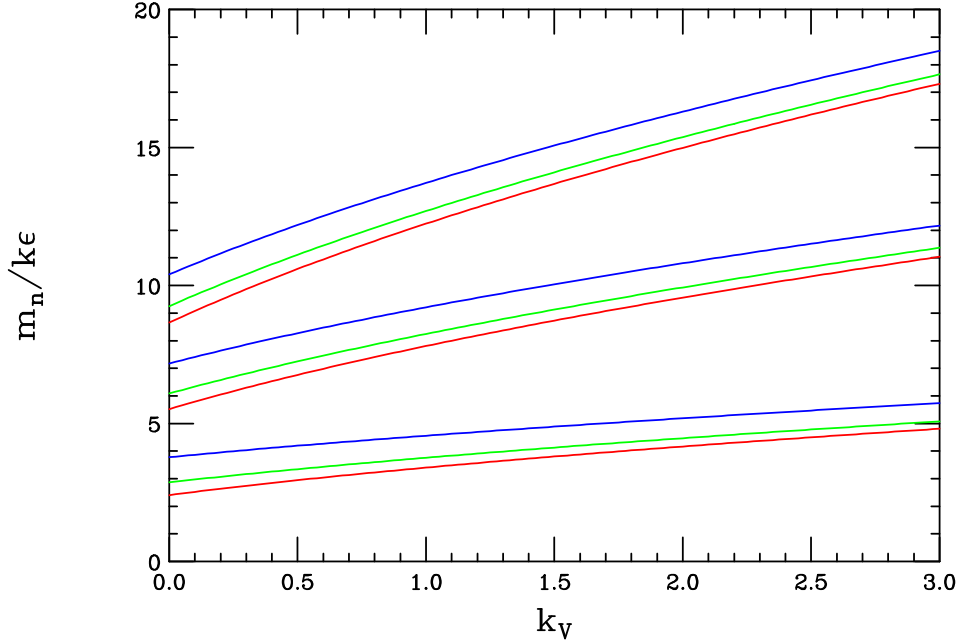}}
\vspace*{0.1cm}
\caption{Masses for the first three massive KK vector excitations as a function of the parameter $k_V$ for, 
from bottom to top, $\mu/k=0,1,2$, respectively.}
\label{fig2}
\end{figure}

Next we turn to the case of bulk gauge fields also allowing for a general non-zero bulk mass term; the action including the 
LIV kinetic term is found to be given by  
\begin{equation}
\int d^4x ~dy ~\sqrt{-g} ~\Bigg[-{1\over {4}} F_{AB}F^{AB}-{k_V \over {4}}\Big(F_{\mu 5}
F^{\mu 5}+F_{5\mu}F^{5\mu}\Big)+{1\over {2}}\mu^2 A_BA^B\Bigg]\,.  
\end{equation}
Here we again see that the LIV leads to a gauge field which effectively `sees' $g_{55}=1+k_V$ which may 
be {\it different} than the effective metric seen by the scalar fields with LIV terms above. 
Following the same path as we did for scalars and taking $A_5=0$ we can easily obtain the 
equation of motion for the gauge field KK wavefunctions, $f_n$, as
\begin{equation}
-(1+k_V)\partial_y \Big[e^{-2ky}\partial_y f_n\Big]+\mu^2e^{-2ky}f_n=m_n^2f_n\,,
\end{equation}
with the solutions 
\begin{equation}
f_n(y)={{e^{ky}}\over {N_n}} \Bigg[J_\nu(\tilde m_n/k ~e^{ky})+ c_{n\nu} Y_\nu(\tilde m_n/k ~e^{ky})
\Bigg]\,,
\end{equation}
where the $c_{n\nu}$ are determined by the boundary conditions as before and the KK masses are given by 
$\tilde m_n^2=m_n^2/(1+k_V)$ where now
\begin{equation}
\nu=\Bigg[1+{{(\mu^2/k^2)}\over {(1+k_V)}}\Bigg]^{1/2}\,,  
\end{equation}
which explicitly depends upon the LIV parameter $k_V$. 
Defining the combination $Z_\nu=J_\nu+c_{n\nu}Y_\nu$, the gauge boson KK masses are given by 
$m_n=k\epsilon x_n\sqrt{1+k_V}$ with the $x_n$ now being the roots of the eigenvalue equation 
\begin{equation}
Z_\nu(x_n)+x_nZ'_\nu(x_n)=0\,.
\end{equation}
Here we again see that we must have $k_V>-1$ in order to prevent the occurrence of tachyonic KK states; 
examples of explicit KK masses and their dependencies on the LIV parameter $k_V$ can be found in 
Fig.~\ref{fig2}. Here we see the same qualitative behavior of the masses as is found for the scalar KK 
case above. The overall general behavior is again observed to be qualitatively similar to that found in the case 
of the flat UED scenario as we see in Fig.~\ref{fig2}. 

\begin{figure}[htbp]
\centerline{
\includegraphics[width=8.5cm,angle=90]{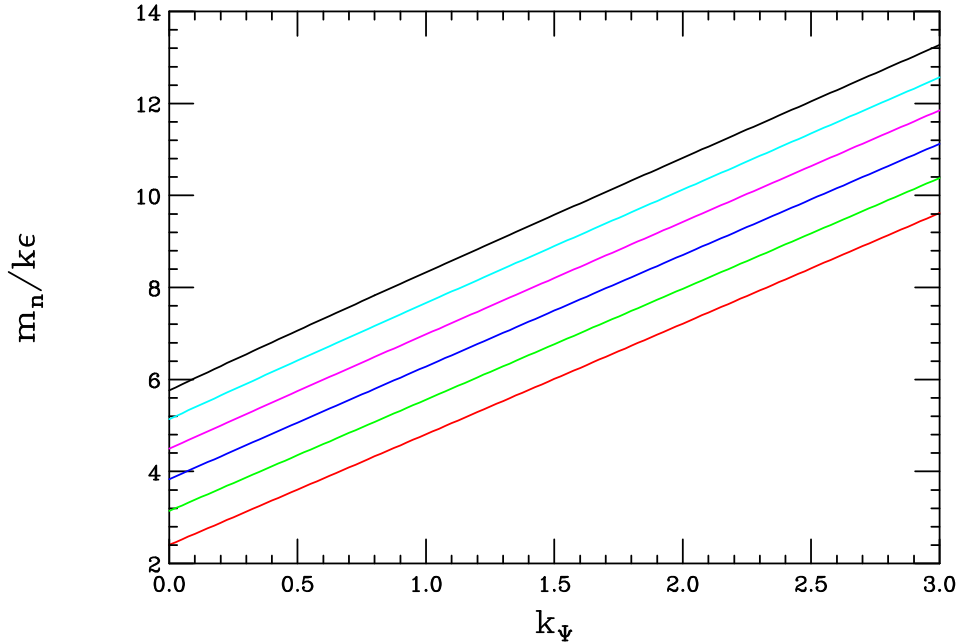}}
\vspace*{0.1cm}
\caption{Masses of the first massive fermion KK excitation as a function of the parameter $k_\Psi$. From bottom to 
top the curves correspond to the values $m_\Psi/k=-1/2, 0,1/2,1,3/2$ and 2, respectively.}
\label{fig3}
\end{figure}

The above analyses for the scalar and vector fields can be immediately extended to the case of bulk 
fermions which are now described by the action 
\begin{equation}
\int d^4x ~dy ~\sqrt {-g}~\Big[V^A_n{i\over {2}}\bar \Psi \gamma^n \bar D_A \Psi-V^5_5{1\over {2}}
k_\Psi \bar \Psi \gamma_5 \bar D_5 \Psi +h.c.-m_\Psi \bar \Psi \Psi\Big]\,,
\end{equation}
with $k_\Psi$ being the fermionic LIV parameter and $V^a_n$ being the appropriate vielbein. Note that in the presence of 
LIV, $\Psi$ effectively feels $g_{55}=1+k_\Psi$. Following the by now standard procedure leads to the 
(almost) familiar coupled pair of first order equations of motion for the KK eigenfunction $g^{L,R}_n$ 
for the LH- and RH-tower modes:
\begin{equation}
(1+k_\Psi)(\pm \partial_y-m_\Psi)g^{L,R}_n=-m_ne^{ky}g^{R,L}_n\\,.
\end{equation}
As before $k_\Psi>-1$ is required to prevent any tachyonic solutions and the eigenfunctions are given by 
\begin{equation}
g^{L,R}_n(y)={{e^{ky/2}}\over {N^{L,R}_n}} \Bigg[J_{1/2\mp\nu}(\tilde m_n/k ~e^{ky})+ \beta_{n\nu} 
Y_{1/2\mp \nu}(\tilde m_n/k ~e^{ky})\Bigg]\,,
\end{equation}
with $\tilde m_n^2=m_n^2/(1+k_\Psi)$ and where $k \nu=m_\Psi/(1+k_\Psi)$. Note that this is a linear 
rescaling of the order of the Bessel functions by the LIV term unlike the quadratic dependence which was 
found for both the bulk scalar and gauge boson cases above. The fermion KK masses are now given by 
$m_n=k\epsilon x_n (1+k_\Psi)$ (again note the linear dependence instead of the square root here) with the 
$x_n$ being the familiar Bessel function roots. An example of the $k_\Psi$ dependence of the lowest fermion 
KK mass for different values of the bulk mass $m_\Psi/k$ is shown in Fig.~\ref{fig3}. Note that unlike the 
scalar and gauge boson cases above the $k_\Psi$ dependence of the KK mass for fermions is nearly linear as 
was also found in the UED analysis for a flat 5-d space.

From the results above it is clear that the scalar, vector and fermion fields which experience LIV in RS-type 
setups respond to this LIV in a qualitatively similar manner as do the corresponding fields in 5-d flat space UED. 
Thus all of the new physics phenomenology that was found in that case which results from the presence of LIV as 
discussed above will also be found in the case of warped extra dimensions though it will differ in numerical detail.

\section{RS Gravity with LIV}

We now turn to the more interesting effects of LIV on the graviton KK spectrum and wave functions in this scenario. 
The existence of LIV in the gravity sector requires us to revisit the setup of the RS model on an 
interval whose original action took the conventional Einstein-Hilbert form:
\begin{equation}
\int d^4x ~dy ~\sqrt{-g}\Bigg[ {{M^3}\over {2}} R- \Lambda \Bigg]\,,  
\end{equation}
where $\Lambda$ is a 5-d cosmological constant. The reason we need to do this is that, amongst other things, the 
presence of the LIV terms in the action will  
lead to modifications of the relationships between the various RS model parameters that we will need to 
incorporate into the equations of motion for the graviton field. Following Kostelecky{\cite {Kost}} (or Carroll 
and Tam{\cite {sean}} for the corresponding flat 5-d case) the simplest possible LIV extended version of the RS 
action takes the form
\begin{equation}
\int d^4x ~dy ~\sqrt{-g} \Bigg[{{M^3}\over {2}} \Big(R+\lambda s^{AB}R_{AB}\Big)-\Lambda\Bigg]\,,  
\end{equation}
with $\lambda$ a dimensionless constant, presumably of order unity, giving the strength of the LIV interaction 
and where in our preferred frame 
$s^{AB}=u^Au^B=\delta^A_5\delta^B_5$, where $u^A=(0,0,0,0,1)$ is the 5-vector discussed above. Here $R_{AB}$ is the 
usual symmetric Ricci tensor. Note that, in principle, an additional term involving the full Riemann curvature 
tensor, \ie, proportional to product 
$\sim u^Au^bu^cu^dR_{ABCD}$, can also be included in the action. However, for our purposes the basic new physics  
associated with this LIV scenario that we wish to examine here can be captured by the simplest case so we will not consider 
the effect of this additional term here except for a few comments below. The equations of motion following from this action 
are found to take the form{\cite {Kost,sean}} of modified Einstein Equations: 
\begin{equation}
G_{AB}=R_{AB}-{1\over {2}}g_{AB}R= {{T_{AB}}\over {M^3}}+\lambda \cal{F_{AB}}\,,  
\end{equation}
where the stress-energy tensor, $T_{AB}=-\Lambda g_{AB}$, is as in the usual RS model and $\cal{F_{AB}}$, which arises 
from the new LIV term in the action, is given by the expression
\begin{equation}
{\cal {F^{AB}}}={1\over {2}}s^{CD}R_{CD}g^{AB}-s^{AD}R_D^B-s^{BD}R_D^A+{1\over {2}}D_C D^A s^{CB}+{1\over {2}}
D_C D^B s^{CA}-{1\over {2}}D^2 s^{AB}-{1\over {2}}G^{AB}D_M D_N s^{MN}\,,  
\end{equation}
where $D_M$ is the covariant derivative on the $AdS_5$ space. Note that while the {\it {partial}} derivative of 
$s^{AB}$ is zero, its covariant derivative is not. To obtain the desired equation of motion for the usual 4-d 
graviton in the transverse, traceless gauge we expand the 4-d part of metric as 
$g_{\mu\nu}=e^{-2ky}(\eta_{\mu\nu}+\kappa h_{\mu\nu})$, keeping only the leading order terms in $\kappa$ and by 
requiring as usual that $h_{55}=h_{\mu 5}=h_\mu^\mu=\partial^\mu h_{\mu\nu}=0$. 

The analysis to obtain the corresponding equation of motion for the KK graviton field from the expression 
above was performed in detail for the case of a {\it flat} 5-d background by Carroll and Tam{\cite {sean}}; these 
authors found the graviton equation of motion 
\begin{equation}
\partial_A\partial^A h_{\mu\nu}= \lambda ~\partial_5^2 h_{\mu\nu}\,,  
\end{equation}
so that the KK masses are given simply by $m_n^2={{n^2}\over {R^2}}(1+\lambda)$ and the wavefunctions take their 
usual trigonometric forms. An important thing 
to notice about their result is that a non-zero value for $\lambda$ does not lead to the generation of a bulk 
mass term in the equation of motion for $h_{\mu\nu}$. This flat space result seems to be completely analogous 
to the those we obtained in the flat UED case for massless 5-d scalars, fermions and gauge fields and is what we 
would perhaps naively imagine the corresponding result to be for gravitons. We will now see the graviton result   
found in the warped case is qualitatively quite different from that which is obtained in flat 5-d space unlike for 
the SM fields.

The $G_{55}$ equation of motion is essentially non-dynamical in the present case and is most easily solved; we 
explicitly obtain 
\begin{equation}
G_{55}=R_{55}-{1\over {2}}R=-6k^2={{\Lambda}\over {M^3}}+2\lambda k^2\,,  
\end{equation}
or more precisely 
\begin{equation}
\Lambda=-6k^2M^3(1+\lambda/3)\,,  
\end{equation}
where the usual RS result is obtained in the $\lambda \to 0$ limit. It is interesting to note from this expression 
that if $\lambda$ becomes large and negative the space undergoes a topological change, $AdS_5 \to dS_5$, which would be inconsistent 
since $\Lambda$ is assumed to be negative. 
Thus in order to maintain the essential physics associated with the $AdS_5$ space of the RS model we must require that $\lambda >-3$. 
We now use this new (\ie, correct) background which satisfies the $G_{55}$ equation of motion above as input and examine the fluctuations around 
the corresponding metric. To this end, inserting the above 
expression for $\Lambda$ back into the Einstein Equations for the $G_{\mu\nu}$ components yields the corresponding 
equation of motion for $h_{\mu\nu}$. We note that once the $G_{55}$ equation of motion above is satisfied (and we use this as an input) 
the `static' part of the $G_{\mu\nu}$ equations of motion on the interval are automatically satisfied since there are no brane tensions to 
tune in this case.  After some lengthy algebra we obtain: 
\begin{equation}
-\partial^\alpha \partial_\alpha h+e^{-2ky}(1+\lambda)\partial_y^2 h-4e^{-2ky}(1+\lambda)k\partial_y h-
8\lambda k^2e^{-2ky}h=0\,,  
\end{equation}
where we have suppressed the indices on $h_{\mu\nu}$ for clarity. Inserting the usual KK decomposition 
\begin{equation}
h_{\mu\nu}=\sum_n h_{\mu\nu}^{(n)}(x) \chi_n(y)\,,  
\end{equation}
and requiring $-\partial^\alpha \partial_\alpha h_{\mu\nu}^{(n)}=m_n^2 h_{\mu\nu}^{(n)}$ as usual, we find that 
the graviton KK wavefunctions must satisfy   
\begin{equation}
-\partial_y \Big[e^{-4ky}\partial_y \chi_n\Big]+m^2e^{-4ky}\chi_n=\tilde {m_n}^2e^{-2ky}\chi_n\,,
\end{equation}
where $\tilde m_n^2=m_n^2/(1+\lambda)$ as we might have expected. If $m^2$ were zero, this result would correspond to 
our naive expectations based on the case of flat 5-d. However, here we observe something new: the non-zero $AdS_5$ 
curvature, $k$, combined with non-zero LIV, \ie,  $\lambda \neq 0$, leads to a bulk mass term for the graviton which is 
given by  
\begin{equation}
m^2={{8\lambda k^2}\over {1+\lambda}}\,,  
\end{equation}
and thus in order to avoid tachyonic KK gravitons we must now require the stronger constraint $\lambda>-1/3$ which can be 
obtained by requiring the index 
$\nu$ in the graviton wavefunctions to be real. From the form of the bulk mass term we observe that in the limit 
that the space becomes flat, $k\to 0$, we immediately recover the result obtained earlier by Carroll and 
Tam{\cite {sean}} in 
flat 5-d space. Clearly the value of the bulk graviton mass that we find when LIV is present depends critically 
upon the nature of the background metric.  

\begin{figure}[htbp]
\centerline{
\includegraphics[width=8.5cm,angle=90]{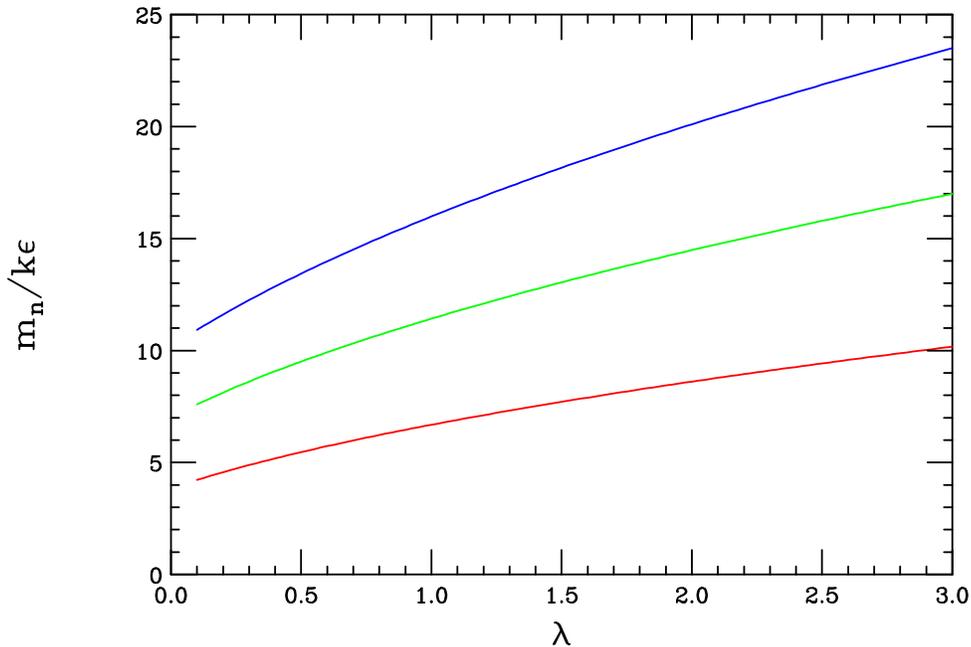}}
\vspace*{0.1cm}
\caption{Masses for the three lightest graviton KK excitations as a function of the parameter $\lambda$.}
\label{fig4}
\end{figure}

As we noted above in the discussion of bulk scalar fields with LIV (which has an equation of motion of the same 
general form), the presence of a bulk mass term in the graviton equation of 
motion prevents the existence of a massless KK zero-mode which we would like to identify with the ordinary massless 
graviton in 4-d. Here we find that the lightest KK mode is {\it massive} due to the breakdown of general co-ordinate 
invariance 
in the 5-d theory that expresses itself here as LIV. Fig.~\ref{fig4} explicitly shows the masses of the three lightest 
graviton KK excitations in this scenario as a function of the parameter $\lambda>0$. Here we explicitly see that the 
lightest KK for $\lambda \neq 0$ is naturally of order the TeV mass scale and that the zero-mode is absent. 

From this analysis we can conclude that LIV in the gravity sector of warped 5-d models leads to a theory which is 
not phenomenologically viable due the generation of a bulk mass term for the 5-d graviton whose existence prevents 
there being a massless graviton in the spectrum. More generally, it would be interesting to know what kind of mass 
terms for gravitons, if any, are generated in other background metrics. It is interesting to speculate that the fact that the metric 
of our ordinary 4-d space depends on the 5th dimension co-ordinate $y$, \ie, the RS metric being non-factorizable while 4-d co-ordinate 
invariance remains unbroken, may be the source of this bulk mass generation effect but this is yet to be demonstrated. A 
general examination of this question would clearly be very interesting but is beyond the scope of the present work. 

Lastly, one can ask what the effect may be of adding the $\sim \tau u^Au^bu^cu^dR_{ABCD}$ term mentioned above where $\tau$ is a constant 
coefficient. If this term alone 
is added we would expect that we would recover an effect similar to that found above although this analysis has not been 
performed. Certainly we would also want to know if the coefficient, $\tau$, could be consistently chosen to maintain the $AdS_5$ 
structure of the RS model while avoiding tachyonic graviton solutions but this is also likely. Having both terms present one could further ask if the 
RS solution could be maintained while simultaneously avoiding the generation of a bulk mass term for the graviton. While we have not 
performed this analysis it is likely that, at best, we may speculate that a fine-tune solution might exist where the values of $\lambda$ and $\tau$ 
could be appropriately chosen to avoid such a graviton bulk mass term. While interesting to contemplate such an analysis is beyond the scope of the 
present paper. 

\section{Summary and Conclusions}

In this paper, extending our earlier work on LIV in UED-type models, we have analyzed the effect of LIV kinetic terms on 
the scalar, gauge, fermion and graviton fields in the 5-d RS model. We found that while the fields of spin-0,1 and 1/2 
responded to LIV in a manner which is qualitatively very similar to that found in the flat 5-d case, something entirely 
new happens in the spin-2 graviton sector. For gravitons, LIV in the curved $AdS_5$ background, unlike in the flat space 
case, leads to the generation of a bulk mass term. The presence of such a term in the KK equations of motion prevents the 
existence of a massless zero-mode corresponding to the 4-d graviton, the lightest field now having a mass in the 
$\sim$ TeV range. Without a massless graviton such a theory cannot be phenomenologically viable. The origin of this 
bulk mass term is directly caused by the loss of general co-ordinate invariance as expressed through spontaneous LIV, 
the non-factorizable nature of the RS metric  and also the non-zero curvature of this background metric about which we are 
expanding the graviton fields. Certainly more detailed analyses of these various possibilities would be both useful and 
interesting to consider.

\noindent{\Large\bf Acknowledgments}

The author would like to thank J.Hewett for discussions related to this work.

\section{Appendix}

One might worry that the substitution of the Eq.(1) into the action Eq.(18) 
does not yield a zero $4-d$ cosmological constant as in the usual RS model 
when the integration over $y$ is performed. It is perhaps simplest to address 
this issue by returning to the usual orbifold boundary condition approach to 
outline the steps that will demonstrate that suitable adjustments can be made 
to the traditional RS solution to obtain this desired result. First, let us 
re-write the metric as 
\begin{equation}
ds^2= e^{-2A}\eta_{\mu\nu}dx^\mu dx^\nu-dy^2\,,
\end{equation}
with $A=k|y|$. The action now takes the form 
\begin{equation}
\int d^4x ~dy ~\sqrt{-g}\Bigg[ {{M^3}\over {2}} (R+\lambda R_{55})- \Lambda -\Lambda_T
\delta(y-\pi r_c)-\Lambda_P \delta(y)\Bigg]\,,
\end{equation}
where $\Lambda_{T(P)}$ are now the tensions on the TeV and Planck branes, 
respectively. Given the form of the metric we might expect that $\Lambda_T=-\Lambda_P$ 
as in the RS case. The above vacuum metric leads to the Ricci tensor elements 
$R_{55}=4[(A')^2-A'']$ and $R_{\mu\nu}=\eta_{\mu\nu}e^{-2A}[A''-4(A')^2]$ 
so that $R=8A''-20(A')^2$. Here, a prime denotes differentiation with respect 
to the co-ordinate $y$. Furthermore, let us recall that $A''=2k[\delta(y)-
\delta(y-\pi r_c)]$. Since the solution to the bulk vacuum equations of motion are 
unaffected by switching to the orbifold approach (the differences being only at the 
brane boundaries) we employ from the solution obtained above that 
$\Lambda=-6k^2M^3(1+\lambda/3)$, which is already linearly shifted from the RS 
value, explicitly in what follows. Substituting all of these results into the above 
action one obtains   
\begin{equation}
\int d^4x ~dy ~e^{-4A}\Bigg[-4k^2M^3(1-\lambda)-\Big(\Lambda_T+kM^3(8-4\lambda)\Big)
\delta(y-\pi r_c)-\Big(\Lambda_P-kM^3(8-4\lambda)\Big) \delta(y)\Bigg]\,.
\end{equation}
Integrating the purely bulk term over $y$ then gives a contribution to the $4-d$ 
cosmological constant of $2kM^3(1-\lambda)(\epsilon^4-1)$ where $\epsilon=e^{-\pi kr_c}$. 
Note that if we also tune the brane tensions by linear shifts in $\lambda$ as 
$\Lambda_T=-\Lambda_P=-(6-2\lambda)kM^3$ then, before the $4-d$ integration,  
the brane contribution to the $4-d$ cosmological constant is found to reduce to
\begin{equation}
\int dy ~e^{-4A} ~2kM^3(1-\lambda)\Bigg[\delta(y)-\delta(y-\pi r_c)\Bigg]= 2kM^3(1-\lambda)(1-
\epsilon^4)\,,
\end{equation}
so that the previously obtained bulk contribution is canceled upon integration and we  
obtain a zero $4-d$ cosmological constant.  
This result indicates that by adjusting the tensions on the TeV and Planck branes, 
as well as the $5-d$ bulk cosmological constant as functions of $k,M$ and $\lambda$ 
a solution can be obtained with a zero $4-d$ cosmological constant even when 
$\lambda \neq 0$. Note that in this approach when the equation of motion for the 
$h_{\mu\nu}$ field is obtained we must apply the usual orbifold boundary conditions at 
$y=0,\pi r_c$. 

Although we have not demonstrated that an identical cancellation occurs in the 
interval approach that we've taken above (by the addition of a Gibbons-Hawking 
surface term) we expect that that this result will also remain valid in that 
case.

%
\def\MPL #1 #2 #3 {Mod. Phys. Lett. {\bf#1},\ #2 (#3)}
\def\NPB #1 #2 #3 {Nucl. Phys. {\bf#1},\ #2 (#3)}
\def\PLB #1 #2 #3 {Phys. Lett. {\bf#1},\ #2 (#3)}
\def\PR #1 #2 #3 {Phys. Rep. {\bf#1},\ #2 (#3)}
\def\PRD #1 #2 #3 {Phys. Rev. {\bf#1},\ #2 (#3)}
\def\PRL #1 #2 #3 {Phys. Rev. Lett. {\bf#1},\ #2 (#3)}
\def\RMP #1 #2 #3 {Rev. Mod. Phys. {\bf#1},\ #2 (#3)}
\def\NIM #1 #2 #3 {Nuc. Inst. Meth. {\bf#1},\ #2 (#3)}
\def\ZPC #1 #2 #3 {Z. Phys. {\bf#1},\ #2 (#3)}
\def\EJPC #1 #2 #3 {E. Phys. J. {\bf#1},\ #2 (#3)}
\def\IJMP #1 #2 #3 {Int. J. Mod. Phys. {\bf#1},\ #2 (#3)}
\def\JHEP #1 #2 #3 {J. High En. Phys. {\bf#1},\ #2 (#3)}


\begin{thebibliography}{99}

\bibitem{livrev}
For some recent overviews, see for example 
  D.~Mattingly,
  Living Rev.\ Rel.\  {\bf 8}, 5 (2005)
  [arXiv:gr-qc/0502097];
  H.~Vucetich,
  arXiv:gr-qc/0502093;
  T.~Jacobson, S.~Liberati and D.~Mattingly,
  Annals Phys.\  {\bf 321}, 150 (2006)
  [arXiv:astro-ph/0505267].

\bibitem{speeds}
  S.~L.~Glashow,
  arXiv:hep-ph/0407087;
  G.~Amelino-Camelia and L.~Smolin,
  Phys.\ Rev.\  D {\bf 80}, 084017 (2009)
  [arXiv:0906.3731 [astro-ph.HE]];
  J.~Bolmont and A.~Jacholkowska,
  arXiv:1007.4954 [astro-ph.HE];
  J.~Ellis, N.~E.~Mavromatos and D.~V.~Nanopoulos,
  Phys.\ Lett.\  B {\bf 674}, 83 (2009)
  [arXiv:0901.4052 [astro-ph.HE]].


\bibitem{dsr}
For a general discussion, see
  G.~Amelino-Camelia,
  Nature {\bf 418}, 34 (2002)
  [arXiv:gr-qc/0207049] and 
  Int.\ J.\ Mod.\ Phys.\  D {\bf 11}, 1643 (2002)
  [arXiv:gr-qc/0210063];
  J.~Kowalski-Glikman,
  arXiv:gr-qc/0603022.


\bibitem{spon}
See, for example, 
  M.~L.~Graesser, A.~Jenkins and M.~B.~Wise,
  Phys.\ Lett.\ B {\bf 613}, 5 (2005)
  [arXiv:hep-th/0501223];
  R.~Bluhm and V.~A.~Kostelecky,
  Phys.\ Rev.\ D {\bf 71}, 065008 (2005)
  [arXiv:hep-th/0412320];
  N.~Arkani-Hamed, H.~C.~Cheng, M.~Luty and J.~Thaler,
  arXiv:hep-ph/0407034; 
  L.~P.~Colatto, A.~L.~A.~Penna and W.~C.~Santos,
  Eur.\ Phys.\ J.\ C {\bf 36}, 79 (2004)
  [arXiv:hep-th/0310220];
  J.~W.~Moffat,
  Int.\ J.\ Mod.\ Phys.\ D {\bf 12}, 1279 (2003)
  [arXiv:hep-th/0211167]; 
  D.~Colladay,
  arXiv:hep-ph/0103021.

\bibitem{Kost}
  D.~Colladay and V.~A.~Kostelecky,
  Phys.\ Rev.\ D {\bf 58}, 116002 (1998)
  [arXiv:hep-ph/9809521] and 
  Phys.\ Rev.\ D {\bf 55}, 6760 (1997)
  [arXiv:hep-ph/9703464];
  V.~A.~Kostelecky,
  Phys.\ Rev.\ D {\bf 69}, 105009 (2004)
  [arXiv:hep-th/0312310];
  V.~A.~Kostelecky and N.~Russell,
  arXiv:0801.0287 [hep-ph].


\bibitem{edrev}
For some recent overviews, see 
  M.~Quiros,
  arXiv:hep-ph/0606153;
  I.~Antoniadis,
  Lect.\ Notes Phys.\  {\bf 720}, 293 (2007)
  [arXiv:hep-ph/0512182];
For an early review, see 
  J.~L.~Hewett and M.~Spiropulu,
  Ann.\ Rev.\ Nucl.\ Part.\ Sci.\  {\bf 52}, 397 (2002)
  [arXiv:hep-ph/0205106].
\

\bibitem{UED}
  T.~Appelquist, H.~C.~Cheng and B.~A.~Dobrescu,
  Phys.\ Rev.\ D {\bf 64}, 035002 (2001)
  [arXiv:hep-ph/0012100];
  H.~C.~Cheng, K.~T.~Matchev and M.~Schmaltz,
  Phys.\ Rev.\ D {\bf 66}, 056006 (2002)
  [arXiv:hep-ph/0205314] and 
  Phys.\ Rev.\ D {\bf 66}, 036005 (2002)
  [arXiv:hep-ph/0204342]; 
  T.~G.~Rizzo,
  Phys.\ Rev.\ D {\bf 64}, 095010 (2001)
  [arXiv:hep-ph/0106336].


\bibitem{flatme}
  T.~G.~Rizzo,
  JHEP {\bf 0509}, 036 (2005)
  [arXiv:hep-ph/0506056].

\bibitem{RS}
  L.~Randall and R.~Sundrum,
  Phys.\ Rev.\ Lett.\  {\bf 83}, 3370 (1999)
  [arXiv:hep-ph/9905221]; 
  H.~Davoudiasl, J.~L.~Hewett and T.~G.~Rizzo,
  Phys.\ Rev.\ Lett.\  {\bf 84}, 2080 (2000)
  [arXiv:hep-ph/9909255], 
  Phys.\ Lett.\  B {\bf 473}, 43 (2000)
  [arXiv:hep-ph/9911262] and 
  Phys.\ Rev.\  D {\bf 63}, 075004 (2001)
  [arXiv:hep-ph/0006041]; 
  A.~Pomarol,
  Phys.\ Lett.\  B {\bf 486}, 153 (2000)
  [arXiv:hep-ph/9911294];
  Y.~Grossman and M.~Neubert,
  Phys.\ Lett.\  B {\bf 474}, 361 (2000)
  [arXiv:hep-ph/9912408];
       T.~Gherghetta and A.~Pomarol,
        Nucl.\ Phys.\  B {\bf 586}, 141 (2000)
        [arXiv:hep-ph/0003129].

\bibitem{nohiggs}
See, for example, 
  C.~Csaki, C.~Grojean, H.~Murayama, L.~Pilo and J.~Terning,
  Phys.\ Rev.\  D {\bf 69}, 055006 (2004)
  [arXiv:hep-ph/0305237].

\bibitem{GW}
  W.~D.~Goldberger and M.~B.~Wise,
  Phys.\ Rev.\  D {\bf 60}, 107505 (1999)
  [arXiv:hep-ph/9907218].

\bibitem{sean}
  S.~M.~Carroll and H.~Tam,
  Phys.\ Rev.\  D {\bf 78}, 044047 (2008)
  [arXiv:0802.0521 [hep-ph]].

\end{thebibliography}
\end{document}